\documentclass[twocolumn,seceq]{jpsj2_cond_mat} 
\usepackage{amssymb}

  \makeatletter
    
    \@addtoreset{equation}{section}
  \makeatother

\title{First-Order Phase Transition with Breaking of Lattice Rotation Symmetry
in Continuous-Spin Model on Triangular Lattice}

\author{Ryo Tamura and Naoki Kawashima}

\inst{Institute for Solid State Physics, University of Tokyo, 5-1-5 Kashiwa-no-ha, Kashiwa, Chiba 277-8581, Japan}

\recdate{\today}

\abst{
Using a Monte Carlo method, we study the finite-temperature phase transition in the two-dimensional classical Heisenberg model on a triangular lattice with or without easy-plane anisotropy.
The model takes account of competing interactions:
a ferromagnetic nearest-neighbor interaction $J_1$ and an antiferromagnetic third nearest-neighbor interaction $J_3$.
As a result,
the ground state is a spiral spin configuration for $-4 < J_1/J_3 < 0$.
In this structure,
global spin rotation cannot compensate for the effect of 120-degree lattice rotation,
in contrast to the conventional 120-degree structure of the nearest-neighbor interaction model.
We find that this model exhibits a first-order phase transition with breaking of the lattice rotation symmetry at a finite temperature.
The transition is characterized as a $Z_2$ vortex dissociation in the isotropic case,
whereas it can be viewed as a $Z$ vortex dissociation in the anisotropic case.
Remarkably,
the latter is continuously connected to the former as the magnitude of anisotropy decreases,
in contrast to the recent work by Misawa and Motome [J. Phys. Soc. Jpn. \textbf{79} (2010) 073001.]
in which both the transitions were found to be continuous.
}

\kword{first-order phase transition, breaking of discrete rotation symmetry, triangular lattice, easy-plane anisotropy, chirality, $Z_2$ vortex, NiGa$_2$S$_4$}

\begin{document}
\maketitle

\newpage


\section{Introduction}

In frustrated magnets, there is no ground state where all interactions are satisfied energetically.
Consequently, conventional magnetic orders are suppressed\cite{diep2005,anderson1973,fazekas1974}.
The frustration appears in antiferromagnets on triangle-based lattices such
as a triangular lattice, a kagome lattice, and a pyrochlore lattice.
It is well-known that interesting ordering phenomena appear in frustrated magnets.
For example, order by disorder on a triangular lattice\cite{nagai1993,lecheminant1995},
as well as on a kagome lattice\cite{reimers1993}, has been reported.
In particular, antiferromagnets that have a triangular lattice are important,
because they represent simplest realization of geometrical frustration.
Thus, the antiferromagnetic model on a triangular lattice has been studied extensively\cite{miyashita1984,kawamura1984_1,kawamura2007,tamura2008,misawa2010}.

The chalcogenide insulator NiGa$_2$S$_4$ has recently been investigated by Nakatsuji and his collaborators\cite{nakatsuji2005,nakatsuji2007_1,nakatsuji2007_2,nakatsuji2010}.
NiGa$_2$S$_4$ has an almost perfectly equilateral triangular lattice that consists of Ni$^{2+}$ ions ($S=1$).
From the susceptibility result, 
the Weiss temperature was estimated to be $\theta_\text{w} = -80$ K, 
which means that strong antiferromagnetic (AF) correlation exists.
This strong AF correlation caused by a third nearest-neighbor (NN) interaction has been suggested from photoemission spectroscopy measurements\cite{takubo2007},
neutron-scattering measurements\cite{stock2010},
and first-principle calculations\cite{mazin2007}.
According to these studies,
the second NN interaction, which is much weaker than the first and third NN interactions,
is negligible.
The magnetic specific heat exhibits an double-peak structure centered near the Weiss temperature $|\theta_\text{w}|$ and 10 K,
and the peaks do not diverge\cite{nakatsuji2005}.
No magnetic long-range order was observed down to 0.35 K in experiments\cite{nakatsuji2005}.
Meanwhile,
short-range incommensurate (IC) correlation, the correlation length of which is a few sites, was detected below 10 K.
Furthermore,
a linearly-dispersive gapless coherent mode in two dimension was observed in this material\cite{nakatsuji2005}.
Thus,
the magnetic short-range order is expected to be strongly related to two-dimensional properties.

Initially,
NiGa$_2$S$_4$ was considered to be an AF material on a triangular (AFT) lattice\cite{collins1997,kawamura1998},
but it has been found that simple AFT Heisenberg models\cite{huse1988,jolicoeur1989} do not describe NiGa$_2$S$_4$ well.
One discrepancy is that the correlation length of spins monotonically develops in the simple AFT models,
whereas it is limited within the range of 10 lattice constants in NiGa$_2$S$_4$.
Another discrepancy is that the magnetic short-range ordered phase does not exist in the simple AFT models.

To explain these discrepancies,
several theoretical models and scenarios have been proposed.
(i) The $S=1$ quantum AF Heisenberg model with NN bilinear-biquadratic interactions on a triangular lattice\cite{tsunetsugu2006,tsunetsugu2007,lauchli2006,bhattacharjee2006} describes the short-range magnetic order,
and the model can explain the gapless linear dispersion observed in NiGa$_2$S$_4$.
Although the quantum bilinear-biquadratic model succeeded in describing a couple aspects of NiGa$_2$S$_4$,
the development of nematic order in the model is not consistent with experimental observations.
(ii) The $Z_2$ vortex scenario has been proposed to describe the phase transition where the specific heat does not diverge.
In the $Z_2$ vortex scenario,
a $Z_2$ topological vortex explains the topological phase transition 
in the classical Heisenberg model with AF NN bilinear interaction on a triangular lattice (AF $J_1$ model)\cite{kawamura1984_1,kawamura1984_2,kawamura2007,kawamura2010}.
However,
an IC correlation has been observed in NiGa$_2$S$_4$,
and the AF $J_1$ model shows a commensurate 120-degree structure.

The AFT model is known to exhibit an IC phase when the model has an AF NN interaction and an AF second NN interaction\cite{jolicoeur1990,chubukov1992}.
Therefore,
since an IC phase appears through competition among several types of interactions,
the classical Heisenberg model on a triangular lattice with NN and third NN bilinear interactions ($J_1$-$J_3$ model) is expected to be a promising model for describing NiGa$_2$S$_4$.
Recently,
we studied the $J_1$-$J_3$ model with the interaction ratio $J_1/J_3 = -1/3$\cite{tamura2008} and found the development of the incommensurate correlation at low temperatures.
We clarified the occurrence of a first-order phase transition with breaking of the lattice rotation symmetry.
However,
this was inconsistent with experimental results for NiGa$_2$S$_4$,
in which a first-order phase transition has not yet been observed.
However,
the $J_1$-$J_3$ model reported in our previous letter suggested an interesting phenomena:
a first-order phase transition caused by discretization of rotation symmetry.
In the previous paper\cite{tamura2008},
we considered the mechanism of the first-order phase transition,
but several points remain to be clarified.

The aim of this paper is to explain the issues listed below.
(i) In the previous study,
a finite-size scaling analysis did not work well because of a correction due to the incommensurability.
Thus,
in this paper,
we adjust the interaction ratio $J_1/J_3$ to realize a commensurate ground state.
This adjustment enables us to eliminate an irrelevant but quite large correction.
In addition,
owing to this adjustment,
we can check whether the incommensurability is essential for the occurrence of the first-order phase transition.
(ii) In the $J_1$-$J_3$ model,
there is the $Z_2$ point defect.
Thus,
we confirm the occurrence of the dissociation of $Z_2$ vortex pairs in the $J_1$-$J_3$ model.
In addition,
we investigate whether the first-order phase transition and the dissociation of $Z_2$ vortex pairs separately occur at finite temperatures.
(iii) In NiGa$_2$S$_4$,
the existence of easy-plane anisotropy has been observed through electron spin resonance experiments\cite{yamaguchi2008,yamaguchi2010}.
Additionally,
in the AF $J_1$ model with easy-plane anisotropy,
different types of phase transitions from the isotropic case have been observed by Capriotti {\it et al}.\cite{capriotti1998,capriotti1999} and Misawa and Motome\cite{misawa2010}.
Thus,
it is interesting to investigate whether the same alteration occurs in the $J_1$-$J_3$ model.
Accordingly,
we examine whether easy-plane anisotropy changes the phase transition in the $J_1$-$J_3$ model.

In \S 2 we introduce the $J_1$-$J_3$ model and investigate its ground state.
We find that the ground state has three types of spiral configurations for $-4<J_1/J_3<0$.
In \S 3 we show the results for the thermal properties in the isotropic $J_1$-$J_3$ model.
To clarify the irrelevance of incommensurability in the first-order phase transition,
we use the interaction ratio to realize a commensurate ground state.
It is clear that incommensurability is not essential for the first-order phase transition,
because the first-order phase transition always takes place.
Furthermore,
to confirm the occurrence of the dissociation of $Z_2$ vortex pairs,
we calculate the temperature dependence of the number density of $Z_2$ vortices.
We find that the dissociation of $Z_2$ vortex pairs occurs at the first-order phase transition temperature.
In \S 4 we discuss the effects of easy-plane anisotropy.
We show the results for the thermal properties of the XY spin limit in the $J_1$-$J_3$ model.
In this case,
we also confirm the existence of the first-order phase transition accompanying the $Z$ vortex dissociation.
From easy-plane anisotropy dependence of the transition temperature and the latent heat,
we find that the $Z$ vortex dissociation is continuously connected to the $Z_2$ vortex dissociation in the $J_1$-$J_3$ model. 
In \S 5 we investigate the interaction ratio dependence of the transition temperature and the latent heat.
From the results,
we summarize the phase diagram of temperature versus interaction ratio and easy-plane anisotropy.
Section 6 is devoted to discussion and summary.


\section{Model and Ground State}

\begin{figure}
\includegraphics[trim=-10mm 0mm 0mm 0mm ,scale=0.40]{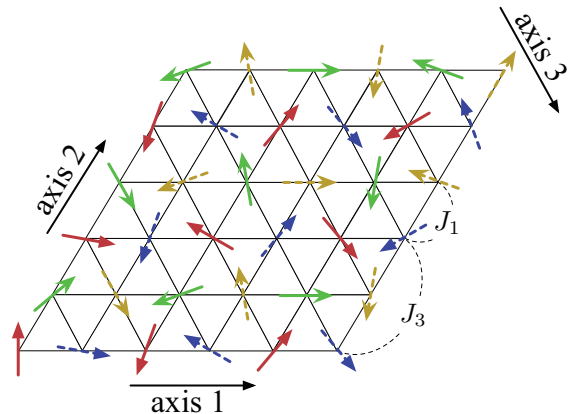} 
\caption{\label{fig:spiral} 
(Color online) Schematic of spin configuration at $\boldsymbol{k}=(k,0)$ and $k=5\pi/9$.
The spins of the same color or type of arrow form a triangular lattice having twice the lattice constant and indicate four sublattices.
}
\end{figure}

The Hamiltonian of the $J_1$-$J_3$ model is given by
\begin{align}
\mathcal{H} = J_1 \sum_{{\langle i , j \rangle}_{\text{NN}}} \boldsymbol{s}_i \cdot \boldsymbol{s}_j +J_3 \sum_{{\langle i , j \rangle}_{\text{3rd. NN}}} \boldsymbol{s}_i \cdot \boldsymbol{s}_j \label{eq:hamiltonian},
\end{align}
where $\boldsymbol{s}_i$ is the vector spin of unit length.
The first sum is taken over NN pairs of sites, and the second sum is taken over third NN pairs (see Fig.~\ref{fig:spiral}).
Here,
we adopt the periodic boundary condition.

In this section,
we derive the ground state spin configuration by mean-field calculation.
The ground state of a classical Heisenberg model is described by 
a spiral configuration with a wave vector $\boldsymbol{k}$ that minimizes the Fourier transform of interactions $J(\boldsymbol{k})$\cite{yoshimori1959,nagamiya1967}.
The spiral configuration is given by
\begin{align}
\boldsymbol{s}_i = \boldsymbol{R} \cos (\boldsymbol{k} \cdot \boldsymbol{r}_i) - \boldsymbol{I} \sin (\boldsymbol{k} \cdot \boldsymbol{r}_i), \label{eq:spiral}
\end{align}
where $\boldsymbol{R}$ and $\boldsymbol{I}$ are two arbitrary orthogonal unit vectors,
and $\boldsymbol{r}_i$ is the position of site $i$ in the real space on a triangular lattice.
Hereinafter,
the lattice constant is set to unity.
We define the primitive translation vectors of a triangular lattice and its reciprocal lattice as $\boldsymbol{a}_1=(1,0)$,  $\boldsymbol{a}_2=(1/2,\sqrt{3}/2)$, $\boldsymbol{b}_1=2 \pi(1,-1/\sqrt{3})$, and $\boldsymbol{b}_2=2 \pi(0,2/\sqrt{3})$, respectively.
If the NN and third NN interactions exist,
the Fourier transform of interactions $J(\boldsymbol{k})$ is given by
\begin{align}
J(\boldsymbol{k}) / J_3 = &J_1/J_3 \left\{ \cos \left( k_x \right) + 2 \cos \left( \frac{1}{2} k_x \right) \cos \left( \frac{\sqrt{3}}{{2}} k_y \right) \right\} \notag \\
&+ \left\{ \cos \left( 2 k_x \right) + 2 \cos \left( k_x \right) \cos \left( \sqrt{3} k_y \right) \right\}.
\end{align}
In this paper,
we consider the case where the third NN interaction is antiferromagnetic ($J_3>0$).
When $J_3$ is positive,
the ground state can be classified into four types depending on the interaction ratio $J_1/J_3$.

\begin{figure}
\includegraphics[trim=-10mm 0mm 0mm 0mm ,scale=0.60]{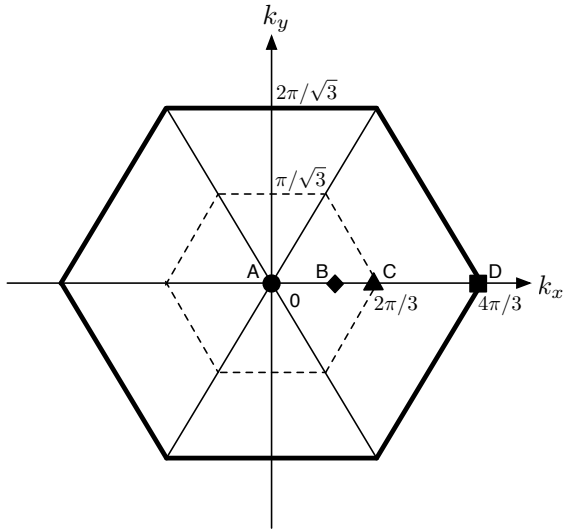} 
\caption{\label{fig:brillouinzone} 
Wave vector $\boldsymbol{k}$ that minimizes $J(\boldsymbol{k})$ in the first Brillouin zone.
Bold hexagon indicates the first Brillouin zone, and its vertices indicate the wave number of the 120-degree structure.
Solid lines between vertices of the first Brillouin zone and center point indicate the lines of 
$\boldsymbol{k}=\pm (k,0)$, $\pm (k/2,\sqrt{3}k/2)$, and $\pm (k/2,-\sqrt{3}k/2)$, respectively.
The vertices of dashed hexagon indicate the wave number of the 60-degree structure.
}
\end{figure}

\begin{enumerate}

\item[(i)] Ferromagnetic ground state ($J_1/J_3 \leq -4$)

In this region, 
the ferromagnetic interaction $J_1$ is dominant, 
and thus $J(\boldsymbol{k})$ has a minimum value at $\boldsymbol{k}=\boldsymbol{0}$ in the first Brillouin zone.
The point at $\boldsymbol{k}=\boldsymbol{0}$ is indicated by point `A' in Fig.~\ref{fig:brillouinzone}.
The order parameter space of the ferromagnetic state is isomorphic to the two-dimensional sphere S$_2$ which corresponds to SO(3)/U(1) symmetry.
The order parameter space is equivalent to the degrees of freedom of the order parameter that characterizes the ground state.

\item[(ii)] Spiral ground state ($-4 < J_1/J_3 < 0$)

In this region, 
neighboring spins in the ground state rotate with pitch governed by the minimum point of $J(\boldsymbol{k})$.
The shift of the minimum point of $J(\boldsymbol{k})$ from $\boldsymbol{k} = \boldsymbol{0}$ is caused by the competition between $J_1$ and $J_3$.
Unlike the AF $J_1$ model, the pitch of the spin rotation is not equal to 120 degrees.
At the six points listed below, $J(\boldsymbol{k})$ takes its minimum value:
\begin{align}
\boldsymbol{k}=\pm (k,0), \pm (k/2,\sqrt{3}k/2), \pm (k/2,-\sqrt{3}k/2). \label{eq:wave_vector}
\end{align}
The wave number $k=|\boldsymbol{k}|$ is given by
\begin{align}
J_1/J_3 = - \frac{2(\sin k + \sin 2 k)}{\sin k + \sin \frac{1}{2} k}, \label{eq:define_int}
\end{align}
where the range of the value of $k$ is $0 < k < 2\pi/3$.
A $k$-point on the $x$-axis that describes a spiral state is indicated by point `B' in Fig.~\ref{fig:brillouinzone}.
The point `B' runs from point `A' to `C' as the value of $J_1/J_3$ increases.
In the case of isotropic Heisenberg spins,
the order parameter space of this spiral state is isomorphic to the three-dimensional real projective space P$_3$ which corresponds to SO(3) symmetry\cite{kawamura1984_1}.
In this case,
the spin configurations characterized by $\boldsymbol{k}$ and $-\boldsymbol{k}$ can be regarded as the same structure.
This is because the difference between the two spin configurations can be eliminated by global SO(3) spin rotation.
Thus,
there are three types of structures, which can be characterized by $\boldsymbol{k}=(k,0)$, $(k/2,\sqrt{3}k/2)$, and $(k/2,-\sqrt{3}k/2)$
at the ground state in the isotropic Heisenberg model.
These three states are separated by energy barriers.
In the spiral spin configuration characterized by these wave vectors, 
there are two types of rotational pitches.
Along one of the three axes, spins rotate with the wave number $k$ (axis 1 in Fig.~\ref{fig:spiral}),
while along the other axes, spins rotate with the wave number $k/2$ (axis 2 and axis 3 in Fig.~\ref{fig:spiral}).
The three types of structures correspond to an axis, characterized by $k$, being selected from among the three axes.
In this structure,
global spin rotation cannot compensate for the effect of 120-degree lattice rotation.
In the spiral ground state region, 
pitches of the spin rotation can be varied by changing the interaction ratio $J_1/J_3$.
In particular, 
a commensurate spiral spin configuration is realized when we set $J_1/J_3$ such that $k/\pi$ is a rational number.

\item[(iii)] Four-sublattice 120-degree structure ($J_1/J_3=0$)

In this parameter case, 
the NN interaction $J_1$ is zero, 
and thus the lattice is divided into four independent sublattices.
Each sublattice is equivalent to the AF $J_1$ model\cite{kawamura1984_1,kawamura1984_2,kawamura2010}.
Thus, 
the ground state is the four-sublattice 120-degree structure.
The minimum points of $J(\boldsymbol{k})$ are $k=2\pi/3$ and $k=4\pi/3$, and these points on the $x$-axis are denoted by points `C' and `D' in Fig.~\ref{fig:brillouinzone}.
The spin configurations characterized by eq.~(\ref{eq:wave_vector}) can be regarded as the same structure.
This is because these structures are not separated by energy barriers in contrast to the case of (ii).
The order parameter space of the four-sublattice 120 degree structure is isomorphic to P$_3$.

\item[(iv)] 120-degree structure ($J_1/J_3>0$)

In this region, 
the NN interaction $J_1$ is antiferromagnetic,
and thus the ground state has the 120-degree structure.
The minimum point of $J(\boldsymbol{k})$ is located at $k=4\pi/3$,
and the point on the $x$-axis is denoted by point `D' in Fig.~\ref{fig:brillouinzone}.
The order parameter space of the 120-degree structure is isomorphic to P$_3$.

\end{enumerate}
The ground states in the $J_1$-$J_3$ model are summarized in Table \ref{tab:groundstate}.
According to this classification,
the spiral ground state, which has three types of structures, only appears in the case of (ii).
Furthermore,
the ferromagnetic phase and 120-degree structure appear in other models and have been thoroughly studied\cite{kapikranian2007,kawamura1984_1}.
Thus,
the interesting parameter range of the $J_1$-$J_3$ model is $-4 < J_1/J_3 < 0$.
In this paper,
we investigate the case where the NN interaction $J_1$ is ferromagnetic and $|J_1|$ is smaller than $|J_3|$.
In \S 3 and \S 4,
we fix the wave number at $k=5\pi/9$.
This value corresponds to $J_1/J_3\cong-0.73425685 \equiv r_{5 \pi/9}$ from eq.~(\ref{eq:define_int}).
The ground state at this parameter is the commensurate configuration.
Along one of the three axes,
the angle between NN spin pairs is 100 degrees (axis 1 in Fig.~\ref{fig:spiral}),
while along the other axes, the angle is 50 degrees (axis 2 and axis 3 in Fig.~\ref{fig:spiral}).


\begin{table*}[t]
\begin{center}
\caption{\label{tab:groundstate} 
Classification of ground states in the isotropic case.
}
\begin{tabular}{c|ccc} \hline
interaction ratio & ground state & number of structure types & type of point defect \\ \hline
$J_1/J_3 \leq -4$ & ferromagnetic state & one & 0 \\
$-4 < J_1/J_3 < 0$ & spiral state & three & $Z_2$ \\
$J_1/J_3=0$(sublattices) & 120-degree structure & one & $Z_2$ \\
$J_1/J_3>0$ & 120-degree structure & one & $Z_2$ \\ \hline
\end{tabular}
\end{center}
\end{table*}


\section{Finite Temperature Properties of Heisenberg Spin Model}


\subsection{First-Order Phase Transition}

\begin{figure}
\includegraphics[trim=-10mm 30mm 0mm 0mm ,scale=0.40]{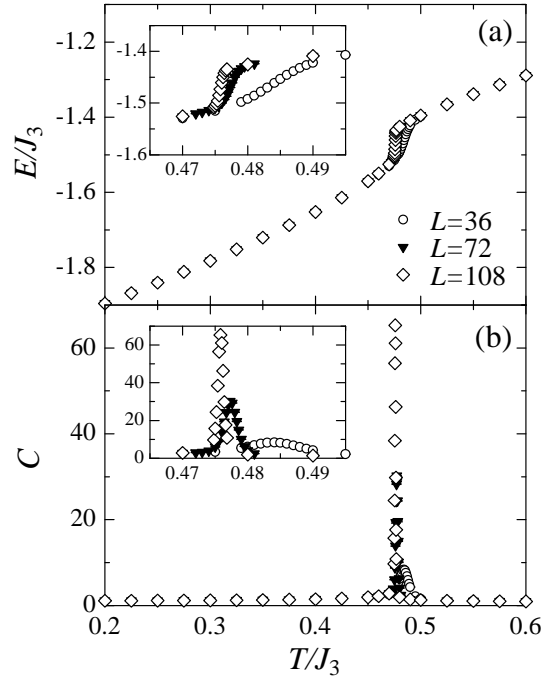} 
\caption{\label{fig:energy_specific} 
Temperature dependence of (a) internal energy per site $E$ and (b) specific heat $C$ for lattice sizes $L=$ 36, 72, and 108.
Insets show enlarged view near the transition temperature.
}
\end{figure}

To discuss a finite-temperature phase transition,
we calculate the internal energy per site $E$ and the specific heat $C$ by Monte Carlo simulation based on the standard heat-bath method.
The specific heat at the temperature $T$ is defined by
\begin{align}
C &= L^2\frac{\langle E^2 \rangle - \langle E \rangle^2}{T^2},
\end{align}
where $\langle \cdots \rangle$ indicates the thermal average.
Hereinafter,
the Boltzmann constant $k_B$ is set to unity.
The interaction ratio $J_1/J_3$ is fixed at $r_{5\pi/9}$ in order that the period of the ground state becomes eighteen sites.
Each simulation contains $10^6$--$10^8$ Monte Carlo steps per spin at each temperature.
We conduct 10--60 independent simulations for each size to evaluate the statistical errors.
We use the conditions below to judge convergence.
\begin{itemize} 
\item[(i)] Physical quantities calculated by short MC steps agree with those calculated by long MC steps.
\item[(ii)] Physical quantities observed in simulations started from random configurations agree with those in simulations started from the ground state.
\end{itemize}
The temperature dependence of the internal energy and the specific heat are shown in Fig.~\ref{fig:energy_specific}.
The specific heat exhibits a single peak which becomes sharp as the lattice size increases.
This behavior indicates the existence of a phase transition in the $J_1$-$J_3$ model.
The peak position moves to lower temperature as the lattice size increases.
The internal energy shows a discontinuous jump around the specific-heat peak.
This behavior is equivalent to that of a first-order phase transition observed at $J_1/J_3=-1/3$
where the ground state is the IC spiral configuration\cite{tamura2008}.

\begin{figure}
\includegraphics[trim=0mm 100mm 0mm 0mm ,scale=0.45]{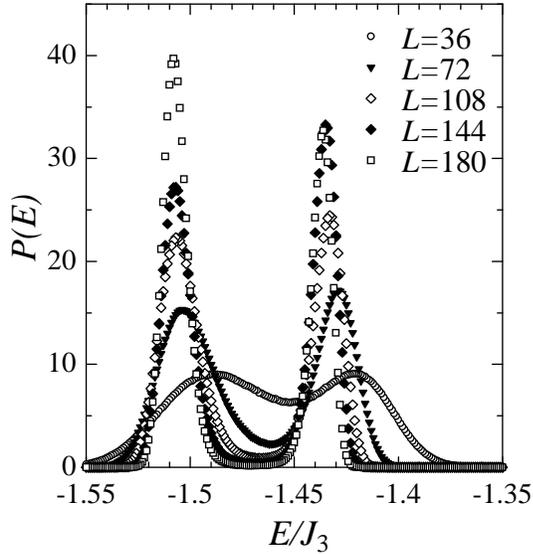} 
\caption{\label{fig:energy_histogram} 
Energy distribution $P(E)$ for
$L$=36 ($T/J_3$=0.4841), $L$=72 ($T/J_3$=0.4773), $L$=108 ($T/J_3$=0.4758), $L$=144 ($T/J_3$=0.4753), and $L$=180 ($T/J_3$=0.4750).
We have confirmed that paramagnetic states and spiral states appear with almost the same probability in eight independent runs.
}
\end{figure}

We investigate the energy distribution to confirm whether the phase transition is of the first order.
Figure~\ref{fig:energy_histogram} shows the energy distribution $P(E)$ near the transition temperature.
A conspicuous bimodal distribution appears near the transition temperature reflecting the same distribution probability of both paramagnetic and spiral states.
We conclude that the phase transition is of the first order.
Even though the ground state is commensurate, 
the $J_1$-$J_3$ model exhibits the first-order phase transition.
Thus, 
the IC magnetic structure is irrelevant to the first-order phase transition in the $J_1$-$J_3$ model.

\begin{figure}
\includegraphics[trim=-4mm 0mm 0mm 0mm ,scale=0.36]{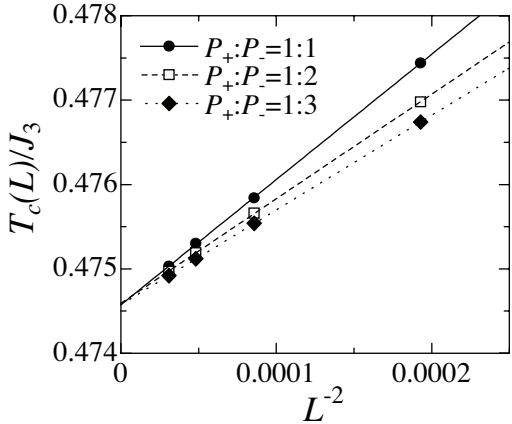} 
\caption{\label{fig:transiton_temperature} 
First-order phase transition temperature plotted as a function of the inverse-square of lattice size.
Straight lines indicate extrapolation of transition temperature.
}
\end{figure}

To estimate the transition temperature $T_c$ for an infinite system,
we adopt the finite-size scaling of the first-order phase transition.
The finite-size effect of $T_c$ is given by
\begin{align}
T_c(L)-T_c \propto L^{-d}, \label{eq:scaling_temperature} 
\end{align}
where $d$ is a dimension of the lattice and $T_c(L)$ is the transition temperature at the lattice size $L$\cite{challa1986}.
We estimate $T_c(L)$ in three ways: 
(i) the ratio of the existence probability of the paramagnetic state $P_+$ and the spiral state $P_-$ is 1:1, 
(ii) the ratio $P_+/P_-$ is 1/2, 
and (iii) the ratio $P_+/P_-$ is 1/3.
The boundary of $P_+$ and $P_-$ is located at the minimum value of $P(E)$ in between the two peaks.
We calculate $P_+$ and $P_-$ by using the reweighting method\cite{ferrenberg1988}.
Reweighting is carried out by using the following relation:
\begin{align}
P(T'; E) \propto e^{-(1/T'-1/T) L^d E} P(T; E),
\end{align}
where $P(T; E)$ is the energy distribution at temperature $T$.
Figure~\ref{fig:transiton_temperature} shows the plot of $T_c(L)/J_3$ versus $L^{-2}$.
The lines in Fig.~\ref{fig:transiton_temperature} are fitting lines calculated by the least-squares method.
The $y$-intercepts of the lines are almost the same.
Therefore,
we obtain the transition temperature at the limit as $L \to \infty$,
\begin{align}
T_c/J_3 = 0.4746(1).
\end{align}
In \S 5,
we discuss the relation between the interaction ratio and the transition temperature.

\begin{figure}
\includegraphics[trim=0mm 140mm 0mm 0mm ,scale=0.45]{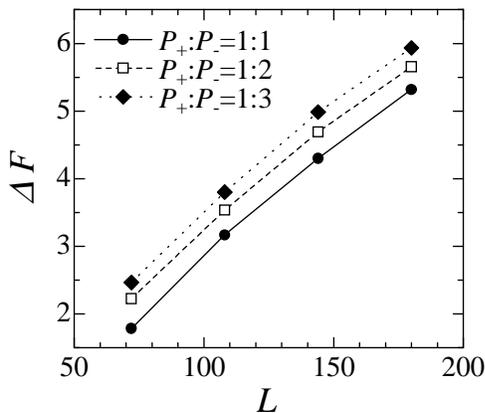} 
\caption{\label{fig:delta_F} 
$\Delta F = \ln(P_\text{max}(E)/P_\text{min}(E))$ plotted as a function of lattice size.
Lines between points are visual guides.
}
\end{figure}

The evidence of the first-order phase transition can be confirmed by other analyses.
We define $P_\text{max}(E)$ and $P_\text{min}(E)$ as the lower-energy peak and the minimum of $P(E)$ between the peaks at $T_c(L)$, respectively.
If the first-order phase transition occurs,
the finite-size scaling of $P_\text{max}(E)$ and $P_\text{min}(E)$ is given by\cite{challa1986,lee1990}
\begin{align}
\Delta F \equiv \ln \left(\frac{P_\text{max}(E)}{P_\text{min}(E)} \right) \propto L^{d-1} \label{eq:scaling_P(E)}.
\end{align}
The size dependence of $\Delta F$ is shown in Fig.~\ref{fig:delta_F}.
The almost linear increase of $\Delta F$ with increasing the lattice size indicates the existence of latent heat at the limit as $L \to \infty$.

\begin{figure}
\includegraphics[trim=-10mm 40mm 0mm 0mm ,scale=0.40]{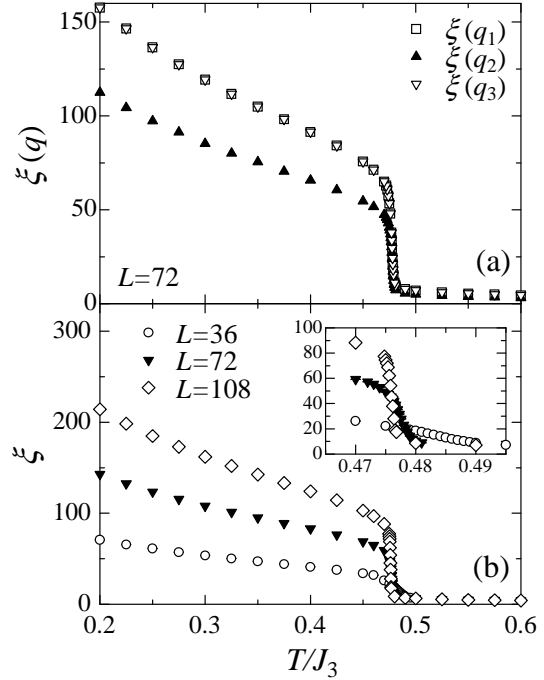} 
\caption{\label{fig:correlation_spin} 
(a) Temperature dependence of the correlation lengths $\xi(\boldsymbol{q}_1)$, $\xi(\boldsymbol{q}_2)$, and $\xi(\boldsymbol{q}_3)$ for $L=72$.
(b) Temperature dependence of the average of three correlation lengths
($\xi = (\xi(\boldsymbol{q}_1) + \xi(\boldsymbol{q}_2) + \xi(\boldsymbol{q}_3))/3$).
Inset shows enlarged view near the transition temperature.
}
\end{figure}

The above discussion about the energy distribution has been shown that the system has the first-order phase transition at a finite temperature.
However, 
the development of the correlation length might be nontrivial.
Because the spin configuration in the spiral state is anisotropic,
the correlation length would exhibit an anisotropic development.
When an anisotropy is strong enough, the critical exponent $\nu$ of correlation length depends on the direction.
In such a case, 
it is known that a usual finite-size scaling (FSS) analysis is inapplicable, 
but a FSS analysis with careful consideration of the anisotropy is required\cite{tonchev2007, binder1989, henkel2002}.
Since our model possibly requires such a special FSS analysis, 
we check the anisotropy of the system by calculating correlation lengths which are dependent on the anisotropy.
Using the Ornstein-Zernike formula\cite{southern1995}, 
correlation lengths are estimated by ratio of structure factor amplitudes.
The structure factor $S(\boldsymbol{q})$ is given by
\begin{align}
S (\boldsymbol{q}) &= \frac{1}{L^2} \sum_{i,j} \langle \boldsymbol{s}_i \cdot \boldsymbol{s}_j \rangle 
e^{i \boldsymbol{q} \cdot (\boldsymbol{r}_i-\boldsymbol{r}_j)},
\end{align}
and the correlation length is estimated by
\begin{align}
\xi (\boldsymbol{q}) = \frac{1}{|\boldsymbol{q}-\boldsymbol{q}_0|} \sqrt{\frac{S(\boldsymbol{q}_0)}{S(\boldsymbol{q})}-1}, \label{eq:correlation_length}
\end{align}
where $\boldsymbol{q}_0$ is one of six $\boldsymbol{k}$'s which satisfy eq.~(\ref{eq:wave_vector}).
When we assume $\boldsymbol{q}_0 = (k, 0)$, the appropriate $\boldsymbol{q}$'s in eq.~(\ref{eq:correlation_length}) are the three vectors of the following:
$\boldsymbol{q}_1 = \boldsymbol{q}_0 \pm \boldsymbol{b}_1/L$,
$\boldsymbol{q}_2 = \boldsymbol{q}_0 \pm \boldsymbol{b}_2/L$, and
$\boldsymbol{q}_3 = \boldsymbol{q}_0 \pm (\boldsymbol{b}_1 + \boldsymbol{b}_2)/L$.
Figure~\ref{fig:correlation_spin}(a), we plot the three correlation lengths $\xi(\boldsymbol{q}_1)$, $\xi(\boldsymbol{q}_2)$, and $\xi(\boldsymbol{q}_3)$ for $L=72$.
Figure~\ref{fig:correlation_spin}(a) clearly shows $\xi(\boldsymbol{q}_2)$ is smaller than $\xi(\boldsymbol{q}_1)$ and
$\xi(\boldsymbol{q}_3)$.
This indicates that fluctuations perpendicular to the axis 1 are larger than those of the perpendicular to axes 2 and 3 (see Fig.~\ref{fig:spiral}).
In order to examine whether the difference in the fluctuations brings about a significant effect on the critical exponent $\nu$,
we compare $\xi(\boldsymbol{q}_2)$ with $\xi(\boldsymbol{q}_1)(= \xi(\boldsymbol{q}_3))$.
The result shows that $\xi(\boldsymbol{q}_2)$ differs only a constant factor from $\xi(\boldsymbol{q}_1)$ ($\xi(\boldsymbol{q}_1)/\xi(\boldsymbol{q}_2) \sim 1.4$),
and thus it is not needed to take care of the anisotropy for a FSS analysis in our spin model.

Figure~\ref{fig:correlation_spin}(b) shows the system size dependence of the correlation length.
To improve the statistical accuracy, we use the average of three correlation lengths
($\xi = (\xi(\boldsymbol{q}_1) + \xi(\boldsymbol{q}_2) + \xi(\boldsymbol{q}_3))/3$).
The every correlation length shows rapid growth at the transition temperature,
and they exceed system sizes in the low-temperature phase.
In this study, 
every simulation shows markedly large correlation lengths below the transition temperature, 
and thus we cannot examine whether the true long-range order exists or not.
However, considering the Mermin-Wagner theorem\cite{mermin1966}, 
we expect that the correlation lengths are finite at finite temperatures even though they are extremely large in the low-temperature phase.


\subsection{Spontaneous Breaking of Threefold Symmetry}

\begin{figure}
\includegraphics[trim=-10mm 30mm 0mm 0mm ,scale=0.40]{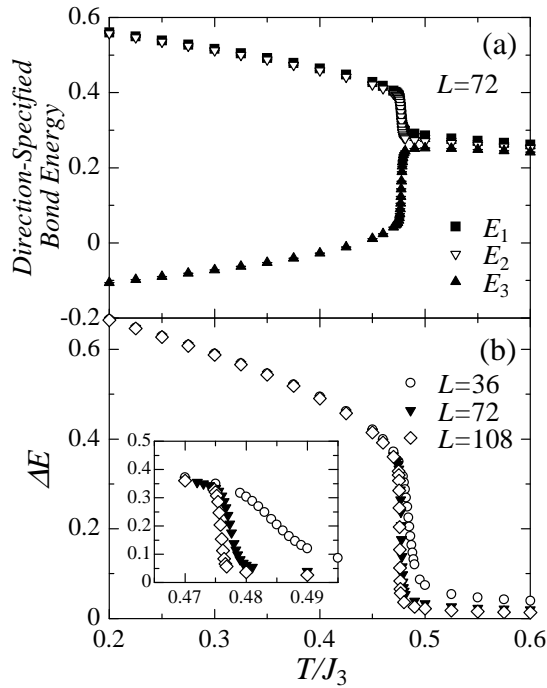} 
\caption{\label{fig:bond_energy}
Temperature dependence of (a) direction-specified bond energies $E_1$, $E_2$, and $E_3$ for $L=72$
and (b) energy difference $\Delta E~(=E_1-E_3)$ for various sizes.
Inset shows enlarged view near transition temperature.
}
\end{figure}

From the ground-state properties of the $J_1$-$J_3$ model,
we expect that the spontaneous breaking of the threefold symmetry occurs at finite temperatures.
At the ground state,
along one of the three axes,
the inner product of NN spin pairs is a negative value characterized by $k$ ($\cos(5\pi/9)\cong-0.17364818$),
while along the other axes, they have a positive value given by $k/2$ ($\cos(5\pi/18)\cong0.64278761$).
Thus,
we calculate $\varepsilon_\mu$ defined by
\begin{align}
\varepsilon_{\mu} &= \frac{1}{L^2} \sum_{{\langle i , j \rangle}_{\text{NN}} \ \parallel \ \text{axis} \ \mu} \boldsymbol{s}_i \cdot \boldsymbol{s}_j, \ \ \ \ (\mu = 1,2,3),
\end{align}
where index $\mu$ indicates one of the three axes of a triangular lattice (i.e., axis 1, 2, or 3 in Fig.~\ref{fig:spiral}).
In the paramagnetic phase,
the values of $\varepsilon_\mu$ do not depend on $\mu$,
because the symmetry does not break.
However,
in the spiral state,
one of the $\varepsilon_\mu$'s is different from the others.
We sort the three averages in descending order and define $E_1$, $E_2$, and $E_3$ as
\begin{align}
E_1 &= \langle \text{max} \{\varepsilon_1,\varepsilon_2,\varepsilon_3 \} \rangle, \notag \\
E_2 &= \langle \text{mid} \{\varepsilon_1,\varepsilon_2,\varepsilon_3 \} \rangle,  \label{eq:bond_energy} \\
E_3 &= \langle \text{min} \{\varepsilon_1,\varepsilon_2,\varepsilon_3 \} \rangle, \notag
\end{align}
where the function ``mid$\{a,b,c\}$'' chooses the second largest value from $a$, $b$, and $c$.  
Since these quantities are the bond energies for the NN pairs along each axis,
we call them the direction-specified bond energies.
Figure.~\ref{fig:bond_energy}(a) shows the temperature dependence of $E_1$, $E_2$, and $E_3$ for $L=72$.
In the paramagnetic phase above $T_c$,
the direction-specified bond energies take the same value, as expected.
While below $T_c$,
$E_1$ and $E_2$ increase but $E_3$ decreases.
This result implies that the threefold symmetry is broken spontaneously at $T_c$.
To study this anomalous behavior quantitatively,
we estimate the energy difference defined by $\Delta E = E_1-E_3$.
The temperature dependence of $\Delta E$ is shown in Fig.~\ref{fig:bond_energy}(b).
The energy difference abruptly increases at $T_c$,
and the gradient of $\Delta E$ appears to diverge as the lattice size increases.
Considering the results,
we conclude that the first-order phase transition accompanies the spontaneous breaking of the threefold lattice rotation symmetry.

Owing to the discrete symmetry breaking,
as in the Potts model\cite{wu1982,tamura2010},
a finite temperature phase transition can occur in the $J_1$-$J_3$ model without violating the Mermin-Wagner theorem.
Recently,
the occurrence of a first-order phase transition with the threefold symmetry breaking
has been reported in other two-dimensional frustrated continuous spin systems\cite{stoudenmire2009,okumura2010}.


\subsection{Relation with $Z_2$ Vortex Dissociation}

Since the order parameter space of 120-degree structure is isomorphic to P$_3$,
the point defect in the AF $J_1$ model is $\Pi_1$(P$_3$)=$Z_2$\cite{kawamura1984_1}.
In the real space,
this point defect corresponds to a vortex configuration labeled by 0 or 1, which is called a $Z_2$ vortex.
It has been reported that the topological transition driven by the dissociation of $Z_2$ vortex pairs occurs in the AF $J_1$ model\cite{kawamura1984_1,kawamura1984_2,kawamura2010}.
The $Z_2$ vortex configuration is created by two unit vectors.
(i) One is the vector chirality defined by
\begin{align}
\boldsymbol{\kappa} (\boldsymbol{r}) = \frac{2}{3 \sqrt{3}} \left( \boldsymbol{s}_1 \times \boldsymbol{s}_2 + \boldsymbol{s}_2 \times \boldsymbol{s}_3 + \boldsymbol{s}_3 \times \boldsymbol{s}_1 \right), \label{eq:vector_chirality}
\end{align}
where the subscript of spin denotes one of the corners on an elementary upward triangle at the position $\boldsymbol{r}$.
(ii) The pointing vector of one spin is located on an elementary upward triangle.
In the simulation,
the $Z_2$ vortex is found by the following procedure.
We calculate the rotation axis $\boldsymbol{n}$ and rotation angle $\omega$ between two orthogonal coordinates defined by two unit vectors.
A variable for the link between two elementary upward triangles labeled $j$ is defined by
\begin{align}
U_j=\exp \left( \frac{\omega_j}{2i} \boldsymbol{n}_j \cdot \boldsymbol{\sigma} \right),
\end{align}
where $\boldsymbol{\sigma}$ is the Pauli matrix.
The vorticity $V [c]$ within a closed contour $c$ is defined by
\begin{align}
V [c] = \frac{1}{2} \left( 1 - \frac{1}{2} \text{Tr} \left( \prod_{j \in c} U_j \right) \right).
\end{align} 
If the contour $c$ involves a $Z_2$ vortex,
$V[c]=1$;
otherwise $V[c]=0$.
Thus,
the number density of $Z_2$ vortices is given by
\begin{align}
n_v = \frac{1}{N_c} \sum_c V [c],
\end{align}
where the sum is taken over all of the smallest closed contours $c$,
and $N_c$ is the number of smallest contours.
This smallest contour is a $\sqrt{3} \times \sqrt{3}$ triangle which is formed by connections between next NN elementary upward triangles on the original triangular lattice.

\begin{figure}
\includegraphics[trim=0mm 140mm 0mm 0mm ,scale=0.40]{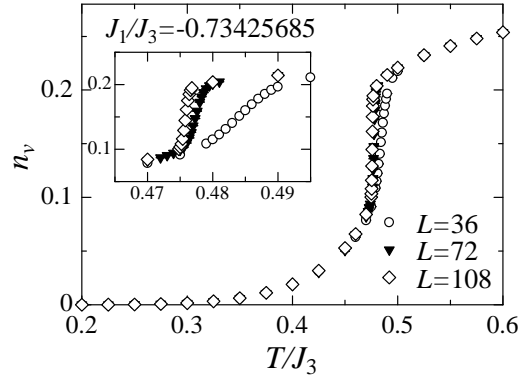} 
\caption{\label{fig:vorticity} 
Temperature dependence of the number density of $Z_2$ vortices $n_v$.
Inset shows enlarged view near the transition temperature.
}
\end{figure}

\begin{figure}
\includegraphics[trim=-15mm 0mm 0mm 0mm ,scale=0.35]{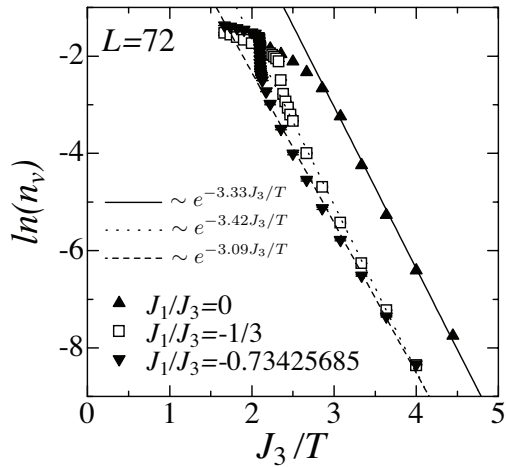} 
\caption{\label{fig:vorticity_arrhenius} 
Arrhenius plot of the number density of $Z_2$ vortices $n_v$ for various interaction ratios when the lattice size is $L=72$.
Lines are the results of the least-squares fitting.
The specific-heat peak corresponding to the first-order phase transition point exists at $J_3/T_c \cong 2.11~(J_1/J_3=r_{5\pi/9})$ and $J_3/T_c \cong 2.38~(J_1/J_3=-1/3)$\cite{tamura2008}.
The topological transition temperature 
in the AF $J_1$ model is $J_3/T_{v} \sim 3~(J_1/J_3=0)$\cite{kawamura1984_1,kawamura2010,southern1993,southern1995,wintel1995}.
}
\end{figure}

The order parameter space of the spiral ground state of the $J_1$-$J_3$ model is isomorphic to P$_3$,
and a $Z_2$ point defect exists.
Thus,
the topological transition may occur in the $J_1$-$J_3$ model, 
as well as in the AF $J_1$ model.
Accordingly,
there is a possibility of a two-step transition:
one is the topological transition caused by a $Z_2$ point defect,
another is the first-order phase transition.
To examine this possibility,
we calculate the number density of $Z_2$ vortices $n_v$.
Since the third NN interaction is dominant in the present model,
we set twice the lattice constant as the unit of length.
Thus,
the limit of $J_1=0$ in the present model is equivalent to the AF $J_1$ model.
Figure~\ref{fig:vorticity} shows the temperature dependence of $n_v$ at $J_1/J_3=r_{5\pi/9}$.
At the transition temperature,
the gradient of $n_v$ appears to diverge as the lattice size increases.
Below the dissociation temperature,
the number density of $Z_2$ vortices obeys the Arrhenius law:
\begin{align}
n_v \propto e^{- 2 \mu /T}, \label{eq:arrhenius}
\end{align}
where $2 \mu$ is the chemical potential of the $Z_2$ vortex pair.
This is because the $Z_2$ vortices are in a bound state below the dissociation temperature.
We show the Arrhenius plot of $n_v$ for $L=72$ in Fig.~\ref{fig:vorticity_arrhenius}
and find that the temperature dependence of $n_v$ fits well to eq.~(\ref{eq:arrhenius}) in the low-temperature phase.
We estimate the dissociation temperature where the data start to violate eq.~(\ref{eq:arrhenius}).
At $J_1/J_3=r_{5\pi/9}$ and $-1/3$,
the dissociation temperature coincides with the first-order phase transition point within the accuracy of the simulation,
and thus we conclude that the dissociation of $Z_2$ vortex pairs occurs at the first-order phase transition temperature.


\section{Effects of Easy-Plane Anisotropy}


\subsection{Model and Ground-State Properties}

In antiferromagnetic systems on a triangular lattice, introduction of easy-plane anisotropy might bring about a change in the behavior of the system.
For example, 
it has been observed that easy-plane anisotropy causes the two separate transitions:
the magnetic Kosterlitz-Thouless transition and the topological $Z_2$ chirality transition~\cite{capriotti1998,capriotti1999,misawa2010}.
Therefore,
it is possible that easy-plane anisotropy changes the behavior of the $J_1$-$J_3$ model.

The model Hamiltonian is given by
\begin{align}
\mathcal{H} = J_1 \sum_{\langle i,j \rangle_{\text{NN}}} \boldsymbol{s}_i \cdot \boldsymbol{s}_j + J_3 \sum_{\langle i,j \rangle_\text{3rd. NN}} \boldsymbol{s}_i \cdot \boldsymbol{s}_j + D_z \sum_i (s_i^z)^2,
\label{eq:hamiltonian_D}
\end{align}
where $s_i^z$ is the $z$ component of spin $\boldsymbol{s}_i$ and $D_z (> 0)$ is easy-plane anisotropy.
We study the ground-state properties for $-4< J_1/J_3 < 0$.
When easy-plane anisotropy is absent ($D_z=0$),
there are three types of structures at the ground state.
Since the ground-state spin configuration is planar,
it is not changed by easy-plane anisotropy. 
However, owing to the $D_z$ term,
all the spins lie in the XY plane at the ground state, and the order parameter space is isomorphic to the one-dimensional sphere S$_1$,
which corresponds to U(1) symmetry.
The spin configurations characterized by wave vector $\boldsymbol{k}$ and $-\boldsymbol{k}$ have different topologies 
in contrast to the isotropic case.
This is because the two spin configurations are different from each other under global U(1) spin rotation.
Thus,
in the anisotropic case,
there are six distinct groups of states corresponding to $\boldsymbol{k}=\pm (k,0)$, $\pm (k/2,\sqrt{3}k/2)$, and $\pm (k/2,-\sqrt{3}k/2)$ at the ground state.
The number of ground states increases in comparison with the isotropic case because of the appearance of topological $Z_2$ chirality.
Furthermore,
the point defect is also different.
The point defect of the $J_1$-$J_3$ model with easy-plane anisotropy is $\Pi_1$(S$_1$)=$Z$, 
whereas the point defect is $\Pi_1(\text{P}_3)=Z_2$ in the isotropic model.
In the following subsections, 
we investigate the lattice rotation symmetry breaking, the $Z_2$ symmetry breaking of chirality, and the dissociation of $Z$ vortex pairs in the anisotropic model.


\subsection{XY Spin Limit}

\begin{figure}
\includegraphics[trim=-10mm 10mm 0mm 0mm ,scale=0.45]{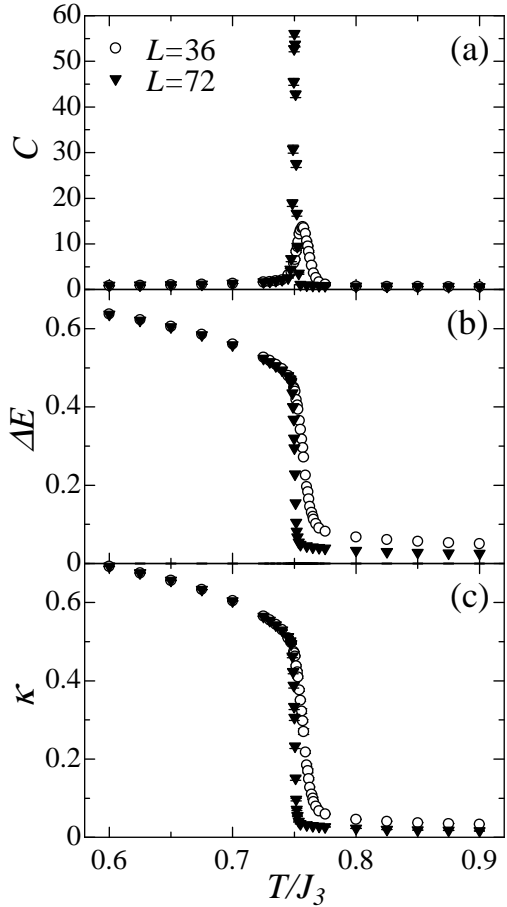} 
\caption{\label{fig:C_deltaE_kappa} 
Temperature dependence of (a) specific heat $C$, (b) $\Delta E~(=E_1-E_3)$, and (c) chirality $\kappa$ of the XY spin limit for $L=36$ and 72.
}
\end{figure}

\begin{figure}
\includegraphics[trim=0mm 100mm 0mm 0mm ,scale=0.45]{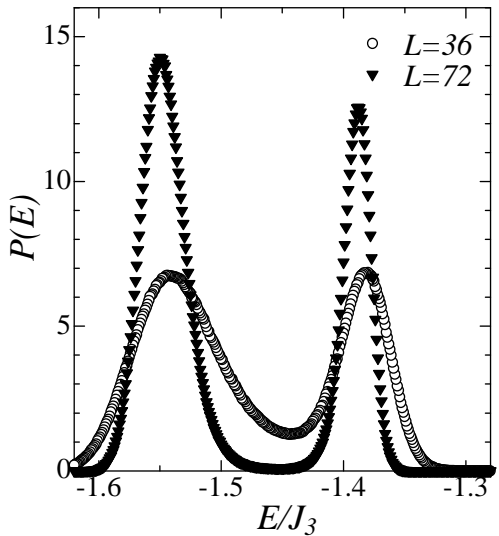} 
\caption{\label{fig:energy_histogram_XY} 
Energy distribution $P(E)$ of the XY spin limit for $L$=36 ($T/J_3$=0.7560) and $L$=72 ($T/J_3$=0.7498).
}
\end{figure}

We investigate finite temperature properties of the XY spin limit ($D_z \to \infty$).
The interaction ratio $J_1/J_3$ is set to $r_{5\pi/9}$, which is the same value as we adopted in \S 3.1.
Whereas we adopted the heat-bath method for the isotropic $J_1$-$J_3$ model, 
we adopt the Metropolis update for the anisotropic model.
Figure~\ref{fig:C_deltaE_kappa}(a) shows the temperature dependence of the specific heat $C$.
The specific heat exhibits a divergent single peak analogous to the isotropic case.
Figure~\ref{fig:energy_histogram_XY} shows the probability distribution of the internal energy $P(E)$ near the temperature at which the peak of $C$ is located.
Although $P(E)$'s for each system size show a bimodal distribution,
the low-energy state seldom undergoes a transition to the high-energy state and \textit{vice versa} when the lattice size is 108.
Thus, 
we omit the data for $L=108$.
Since the valley in the middle of $P(E)$ of $L=72$ is obviously deeper than that of $L=36$, 
we conclude that the phase transition is of the first order.

To investigate the spontaneous symmetry breaking at the transition temperature,
we calculate the energy difference of the direction specified bond energies $\Delta E$ and the chirality $\kappa$.
The chirality defined by eq.~(\ref{eq:vector_chirality}) becomes scalar for anisotropic spins.
$\Delta E$ and $\kappa$ are plotted versus temperature in Figs.~\ref{fig:C_deltaE_kappa}(b) and \ref{fig:C_deltaE_kappa}(c), respectively.
These quantities abruptly increase at the first-order phase transition temperature.
$\Delta E$ and $\kappa$ correspond to the order parameters of lattice rotation symmetry breaking and $Z_2$ symmetry breaking of chirality, respectively.
Therefore,
we conclude that the first-order phase transition accompanies the spontaneous breaking of the sixfold symmetry.

\begin{figure}
\includegraphics[trim=-20mm 0mm 0mm 0mm ,scale=0.35]{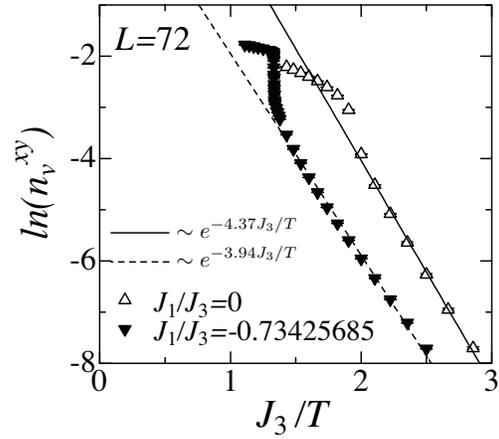} 
\caption{\label{fig:vorticity_arrheniusXY} 
Arrhenius plot of the number density of $Z$ vortices $n_v^{xy}$ of the XY spin limit for $L=72$.
Lines are the results of least-squares fitting.
The specific-heat peak corresponding to the first-order phase transition point exists at $J_3/T_c \cong 1.33~(J_1/J_3=r_{5\pi/9})$.
The dissociation temperature of $Z$ vortex pairs
in the XY spin version of the AF $J_1$ model is $J_3/T_v \sim 2~(J_1/J_3=0)$\cite{miyashita1984}.
}
\end{figure}

To study the dissociation of $Z$ vortex pairs,
we calculate the number density of $Z$ vortices $n_v^{xy}$.
The number density of $Z$ vortices is calculated in the same manner as in ref.~\ref{ref:tobochnik}.
The Arrhenius plot of $n_v^{xy}$ for $L=72$ is shown in Fig.~\ref{fig:vorticity_arrheniusXY}.
The data are well fitted by the Arrhenius law in the low-temperature phase.
We estimate the dissociation temperature as described in the previous section.
In Fig.~\ref{fig:vorticity_arrheniusXY},
the dissociation temperature appears to coincide with the first-order phase transition temperature.
The Arrhenius plot indicates that $Z$ vortex pairs dissociate at the first-order phase transition temperature within the accuracy of the simulation.
From these results, 
the first-order phase transition in the XY spin limit of the $J_1$-$J_3$ model accompanies
the lattice rotation symmetry breaking,
the $Z_2$ symmetry breaking of chirality,
and the dissociation of $Z$ vortex pairs.


\subsection{Finite Easy-Plane Anisotropy}

\begin{figure}
\includegraphics[trim=-10mm 30mm 0mm 0mm ,scale=0.40]{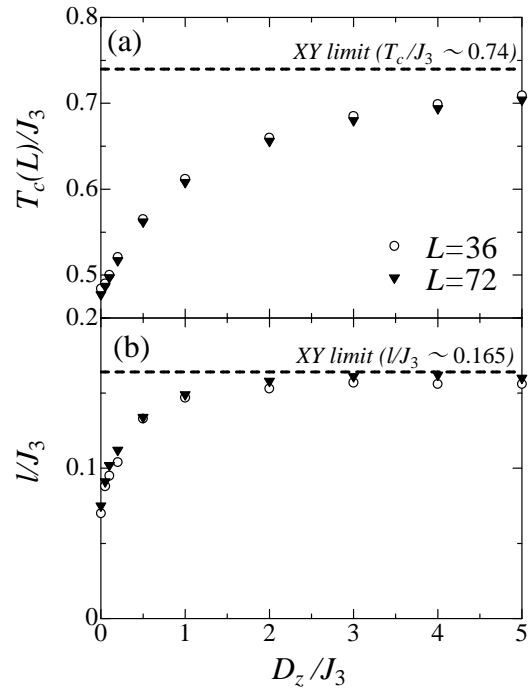} 
\caption{\label{fig:spin_anisotropy} 
Easy-plane anisotropy $D_z$ dependence of (a) first-order phase transition temperature $T_c(L)$ and (b) latent heat $l$.
The point where $D_z=0$ corresponds to the isotropic model.
The dashed lines indicate the results for the XY spin limit ($D_z \to \infty$). 
}
\end{figure}

In this subsection, 
we consider a finite easy-plane anisotropy ($D_z>0$).
Figure~\ref{fig:spin_anisotropy} shows the transition temperature $T_c(L)$ and the latent heat $l$ as a function of lattice size $L$.
The transition temperature $T_c(L)$ is defined as the point where the bottom of the valley in $P(E)$ divides $P(E)$ equally. 
The latent heat $l$ is estimated from the width of the bimodal distribution.
Both $T_c(L)$ and $l$ continuously increase as the anisotropy $D_z$ increases.
In other words, 
although the introduction of $D_z$ raises the transition temperature and the latent heat, 
it does not change the essential properties of the $J_1$-$J_3$ model.
Therefore,
the $Z$ vortex dissociation is continuously connected to the $Z_2$ vortex dissociation as the magnitude of anisotropy decreases in the $J_1$-$J_3$ model.

\section{Dependence of Interaction Ratio}

\begin{figure}
\includegraphics[trim=-10mm 30mm 0mm 0mm ,scale=0.40]{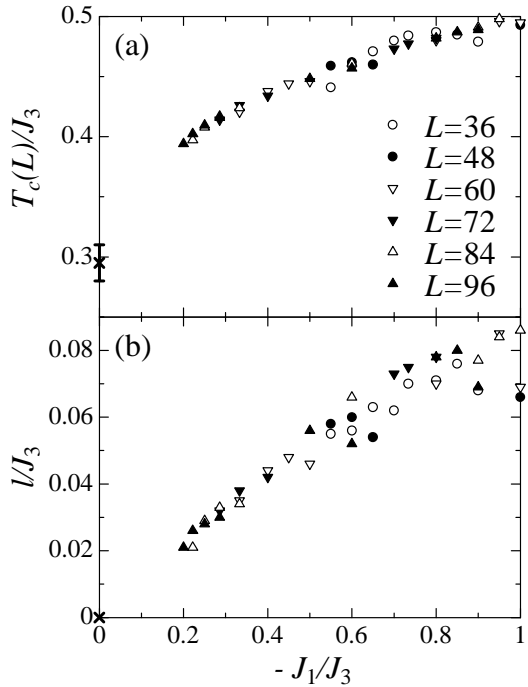} 
\caption{\label{fig:various_interaction} 
(a) First-order phase transition temperature $T_c(L)$ and (b) latent heat $l$ as a function of $- J_1/J_3$
for lattice sizes $L=$ 36, 48, 60, 72, 84, and 96.
The cross points at $J_1/J_3=0$ indicate the temperature of the dissociation of $Z_2$ vortex pairs $T_{v}/J_3 \sim 0.3$\cite{kawamura1984_1,kawamura2010,southern1993,southern1995,wintel1995} and the continuous transition ($l=0$).
}
\end{figure}

\begin{figure}
\includegraphics[trim=3mm 140mm 0mm 0mm ,scale=0.44]{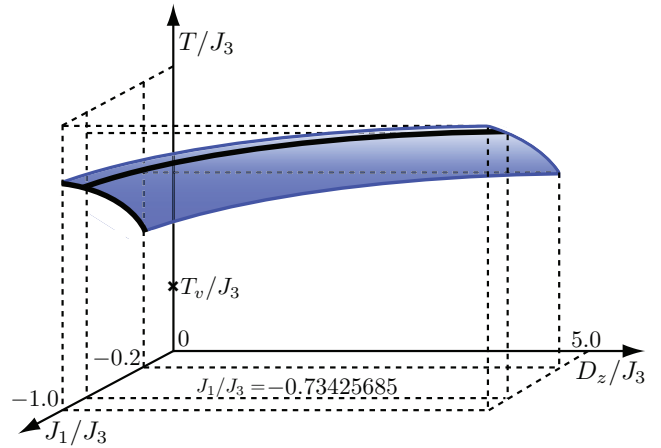} 
\caption{\label{fig:phase_diagram_3D}
(Color online)
Schematic phase diagram of temperature $T$ versus interaction ratio $J_1/J_3$ and easy-plane anisotropy $D_z$. 
Bold lines indicate the boundaries of first-order phase transition found in this work.
Blue (light-colored) plane indicates the predicted boundary of the first-order phase transition.
}
\end{figure}

So far, 
we have considered the first-order phase transitions of the $J_1$-$J_3$ model.
However, 
when we set $J_1$ to be zero, 
the system does not show the specific-heat anomaly\cite{kawamura1984_1}.
Thus, 
there should be a threshold value of $J_1$ where the phase transition changes from being continuous to being discontinuous.
In this section, 
we examine the interaction ratio dependence of the isotropic Heisenberg model.
Since the period of the ground-state spin configuration becomes quite large for the small $J_1$ system, 
we restrict the range of $J_1/J_3$ to $-1.0 \leq J_1/J_3 \leq -0.2$.
To examine whether the phase transition is continuous, 
we calculate the probability distribution of the internal energy.
In the range of $J_1/J_3$ used here, 
the probability distribution is always bimodal near the critical point, 
and thus the phase transitions are of the first order for $-1.0 \leq J_1/J_3 \leq -0.2$.
Figure \ref{fig:various_interaction} shows the first-order phase transition temperature $T_c(L)$ and the latent heat $l$ as functions of $J_1/J_3$.
The phase diagrams show that the transition temperature continuously increases as the interaction ratio $|J_1/J_3|$ increases,
and the latent heat is also proportional to $-J_1/J_3$.
In the simulation, 
we observe neither a new phase boundary nor a threshold value of $J_1$ where the phase transition changes from being continuous to being discontinuous.
Thus,
at least for $J_1/J_3$ to $-1.0 \leq J_1/J_3 \leq -0.2$,
the phase boundary exists only between the paramagnetic phase and the spiral phase,
and the phase transition is of the first order.
However, 
we cannot exclude the possibility that the phase transition is continuous below a small $J_1$.
We summarize the phase diagram of temperature versus interaction ratio and easy-plane anisotropy in Fig.~\ref{fig:phase_diagram_3D}.
Figure \ref{fig:phase_diagram_3D} shows that the first-order phase transition is quite common for various environments in the $J_1$-$J_3$ model.

\section{Discussion and Summary}

We discuss the implications of the present results in regard to the experimental results for NiGa$_2$S$_4$.
In the $J_1$-$J_3$ model,
even if the value of $J_1/J_3$ changes and easy-plane anisotropy is added,
the thermal phase transition is of the first order.
This is inconsistent with experimental results for NiGa$_2$S$_4$,
because clear evidence of a first-order phase transition has not been observed.
Furthermore,
the spin correlation estimated from neutron diffraction measurements reached only a few sites at low temperatures,
in contrast with the large correlation length estimated by simulation.
Therefore,
the low-temperature phase of the $J_1$-$J_3$ model are qualitatively different from that observed in NiGa$_2$S$_4$.

In this paper, 
we have studied the critical phenomena of the $J_1$-$J_3$ model on a triangular lattice by a Monte Carlo method.
We have set the nearest-neighbor interaction $J_1$ and the third nearest-neighbor interaction $J_3$ to be ferromagnetic and antiferromagnetic, respectively.
When the interaction ratio $J_1/J_3$ is set to be $-4 < J_1/J_3 < 0$, 
the spin configuration of the ground state forms a spiral.
In this spiral configuration,
global spin rotation cannot compensate for the effect of 120-degree lattice rotation,
and there are three types of distinct structures.
We have considered the phase transition of the $J_1$-$J_3$ model under three conditions.
(i) The interaction ratio $J_1/J_3$ is set to $-0.73425685$ such that the ground state is commensurate. Adopted spins are isotropic.
(ii) An extra anisotropic term, $D_z \sum_i (s_i^z)^2$, is added to the Hamiltonian.
(iii) Several values of the interaction ratio $J_1/J_3$ are incorporated into the isotropic model.

In \S 3,
under condition (i), 
we have investigated whether the incommensurate nature is essential.
Since a clear first-order phase transition is seen in the system where the ground state is commensurate,
we conclude that incommensurability is irrelevant to the discontinuous phase transition.
We have calculated the energy difference $\Delta E$ and the number density of $Z_2$ vortices $n_v$ near the transition point.
The energy difference $\Delta E$ is an indicator of threefold symmetry breaking, 
and the number density of $Z_2$ vortices $n_v$ follows the Arrhenius law below the dissociation temperature.
Since both $\Delta E$ and $n_v$ exhibit the anomalous behavior at the same time, 
the breaking of threefold lattice rotation symmetry and the dissociation of $Z_2$ vortex pairs occur at the same first-order phase transition point.

In \S 4, 
we have considered the effects of easy-plane anisotropy on the first-order phase transition under condition (ii).
In this case,
we have also found the existence of the first-order phase transition, as in the isotropic case.
Even when easy-plane anisotropy is introduced into the model,
the order of the thermal transition does not change.
Furthermore,
this first-order phase transition accompanies the lattice rotation symmetry breaking,
the $Z_2$ symmetry breaking of chirality,
and the dissociation of $Z$ vortex pairs.
Therefore,
the essential properties of the $J_1$-$J_3$ model are not change by taken easy-plane anisotropy into account.
In other words,
the $Z$ vortex dissociation is continuously connected to the $Z_2$ vortex dissociation in the $J_1$-$J_3$ model.

In \S 5, 
under the condition (iii), we have examined whether lowering the interaction ratio would bring about a continuous phase transition.
We have found that a first-order phase transition occurs for $-1.0 \le J_1/J_3 \le -0.2$ in the isotropic case.
However,
the possibility of the continuous phase transition still remains for $-0.2 \le J_1/J_3$.

In spite of the fact that we have adopted several conditions, as mentioned above,
the $J_1$-$J_3$ model always exhibits a first-order phase transition.
From this finding,
the first-order phase transition in the $J_1$-$J_3$ model is quite common and robust against changing environments.
Therefore,
we expect that such a first-order phase transition will be observed in actual frustrated magnets with competing interactions.


\section*{Acknowledgments}

We would like to thank Shu Tanaka, Takahumi Suzuki, Yusuke Tomita, Satoru Nakatsuji, Hirokazu Tsunetsugu, and Hikaru Kawamura for useful comments and discussions. 
R.T. acknowledges support from the Global COE Program ``The Physical Sciences Frontier'' of the Ministry of Education, Culture, Sports, Science and Technology (MEXT), Japan.
This work is financially supported by 
Grants-in-Aid for Scientific Research (B) (22340111) and for Scientific Research on Priority Areas ``Novel States of Matter Induced by Frustration'' (19052004) from MEXT, Japan, 
as well as by Next Generation Supercomputing Project, Nanoscience Program, MEXT, Japan.
The computation in this work is executed on computers at the Supercomputer Center, Institute for Solid State Physics, University of Tokyo.


\end{document}